\documentclass[10pt,preprint]{aastex}

\begin{document}

\title{On the Near--Infrared Size of Vega}

\author{David R. Ciardi}
\affil{211 Space Sciences Building, Department of Astronomy, University of 
Florida, Gainesville, FL 32611}
\email{ciardi@astro.ufl.edu}

\author{Gerard T. van Belle}
\affil{Jet Propulsion Laboratory, California Institute of Technology,
MS 171-113, 4800 Oak Grove, Pasadena, CA 91109}
\email{gerard@huey.jpl.nasa.gov}

\author{Rachel L. Akeson\altaffilmark{1}}
\affil{Infrared Processing and Analysis Center, California Institute of 
Technology MS 100-22, 770 South Wilson Avenue, Pasadena, CA, 91125}
\altaffiltext{1}{Jet Propulsion Laboratory, California Institute of 
Technology, MS 171-113, 4800 Oak Grove, Pasadena, CA 91109}
\email{rla@ipac.caltech.edu}

\author{Robert R. Thompson\altaffilmark{2}}
\affil{Jet Propulsion Laboratory, California Institute of Technology,
MS 171-113, 4800 Oak Grove, Pasadena, CA 91109} 
\altaffiltext{2}{Department of Physics \& Astronomy, University of 
Wyoming, Laramie, WY 82071-3905}
\email{thompson@huey.jpl.nasa.gov}

\author{Elizabeth A. Lada}
\affil{211 Space Sciences Building, Department of Astronomy, University of 
Florida, Gainesville, FL 32611}
\email{lada@astro.ufl.edu}

\and

\author{Steve B. Howell}
\affil{Astrophysics Group, Planetary Science Institute, Tucson, AZ 85705}
\email{howell@psi.edu}

\slugcomment{Accepted by The Astrophysical Journal}

\begin{abstract}

Near--infrared (2.2 \micron) long baseline interferometric observations
of Vega are presented. The stellar disk of the star has been resolved,
and the data have been fitted with a limb darkened stellar disk of
diameter $\Theta_{LD} = 3.28 \pm 0.01$ mas. The derived effective
temperature is $T_{eff} = 9553 \pm 111$ K. However, the residuals
resulting from the stellar disk model appear to be significant and
display organized structure. Instrumental artifacts, stellar surface
structure, stellar atmosphere structure, and extended
emission/scattering from the debris disk are discussed as possible
sources of the residuals . While the current dataset cannot uniquely
determine the origin of the residuals, the debris disk is found to be
the most likely source. A simple debris disk model, with 3-6\% of Vega's
flux emanating from the disk at $r \lesssim 4$ AU, can explain the
residuals.

\end{abstract}


\section{Introduction}

Vega ($\alpha$ Lyrae=HD 172167=HR 7001, A0V, d=7.76 pc) is arguably one
of the most important stars in astrophysics. It has been used
extensively as a spectrophotometric absolute flux standard in the
optical, ultraviolet \citep{hayes85, bohlin90}, and infrared
\citep{cohen92}. Vega is one of the few main sequence stars for which an
angular diameter measurement has been made ($\Theta_{LD} = 3.24 \pm
0.07$; \citet{hbda74}), and Vega has been used as a template for our
fundamental understanding and modeling of stellar atmospheres
\citep[e.g.,][] {kurucz79, db80, ck94}.

In addition to the importance of Vega for stellar astrophysics, Vega has
been fundamental to our understanding of exo-zodiacal and extra-solar
planetary systems. \citet{aumann84} reported that {\it IRAS} had
detected an infrared excess above what was expected for the stellar
photosphere at $ \lambda \gtrsim 12$ \micron. The infrared excess has
been attributed to a circumstellar disk believed to be a denser analog
of our own zodiacal cloud.

Many questions regarding the debris disks around other stars remain
unanswered. How much total mass is located within the disks? What is the
morphology of the disks? What is the size distribution of the dust
grains? What is the composition of the dust? What is the lifetime of the
disks? Are the debris disks leftovers of active planet formation?

The intensity contrast between the stellar photosphere and the disk
makes observing the disk at short wavelengths ($\lambda \lesssim 10$
\micron) difficult. As a result, most of the work on the debris disk
properties of Vega (and other Vega-like sources) has been performed at
far-infrared, submillimeter, and millimeter wavelengths. These
observations have coarse spatial resolution, and the wavelengths are
relatively insensitive to smaller, hotter grains located close to the
star. Consequently, the data are biased to cooler and larger grains
located in the outer portions of the debris disk ($r \gtrsim 50-100$
AU), where the equilibrium temperature is $T \lesssim 100$ K
\citep{bp93}, and comparatively little is known about the inner regions
of debris disks ($r \lesssim 10$ AU).

Dust grains located within a few AU of the star are expected to
contribute most significantly at $\lambda \lesssim 10$ \micron, either
through emission or scattering. Some Vega-like stars do indeed show
near-infrared excesses with hot dust ($T \sim 500 - 1500$ K) located
within a few AU of the star \citep[e.g.,][]{ssb97}. Vega-like sources
with near-infrared excesses may be younger than sources that exhibit
only longer wavelength ($\lambda \gtrsim 10$ \micron) excesses.

Previous work is inconclusive regarding the presence of a near-infrared
excess associated with Vega. Stellar atmosphere models have consistently
underestimated the measured infrared ($\lambda \gtrsim 2$ \micron) flux
of Vega \citep{mountain85, leggett86a}. However, \citet{leggett86b}
compared the near-infrared colors of Vega with 25 B and A stars and
found the colors of Vega to be consistent with these stars within the
uncertainties (a few percent). This led \citet{leggett86b} to suggest
that perhaps stellar atmosphere models are incorrect and that the models
underestimate the photospheric infrared flux of A-type stars.
\citet{bcp98} comment it is difficult to understand where the current
stellar atmosphere models for A-type stars fail, as the dominating H
opacity in A stars is thought to be very well understood. Finally,
\citet{megessier95} reviews the optical and near-infrared calibrations
of Vega up through 1995 and critically examines the various calibration
methods. \citet{megessier95} finds that Vega is apparently brighter in
the near-infrared than a normal A0V star by $\sim 0.04 $ mag (3.5\%) at
K, with an increasing excess as the wavelength increases. Thus, as well
studied as Vega is, it is unclear whether or not it has an excess at
$\lambda \lesssim 10$ \micron.

If the disk is detectable, either as emission or scattering, at $\lambda
\lesssim 10$ \micron, this would have important implications regarding
the dust temperature, size distribution, composition, and overall
morphology of the debris disk surrounding Vega, and Vega-like stars, in
general. Unfortunately, the apparent near-infrared excess relative to
current stellar atmosphere models is of the same order of magnitude as
the flux calibration uncertainties (2-4\%; \citet{leggett86a}), and it
is difficult to determine from photometry alone if Vega truly has a
near-infrared excess.

In this paper, the possible near-infrared excess associated with Vega
has been investigated via another method independent of photometry:
infrared interferometry. Vega was observed over two seasons with the
Palomar Testbed Interferometer (PTI, \citet{colavita99}), in an attempt
to ascertain if the debris disk around Vega is detectable by PTI at 2.2
\micron. The interferometric experiment does do not rely upon precise
infrared photometry, but rather on the capability of the interferometer
to detect spatial structure (i.e., whether or not there is any extended
emission beyond the stellar photosphere).

The details of the observations and data reduction are discussed in \S
2, and in \S 3, the results and analysis are discussed. Overall, the
observed visibility curve is dominated by the resolved stellar disk (as
expected). However, after modeling the data with a stellar photospheric
disk, significant residuals, which display organized structure, remain.
The possible sources of the residuals, including instrumental artifacts,
stellar surface structure, stellar atmospheric structure, and extended
emission/scattering from the debris disk, are explored.

\section{Observations and Data Reduction}

PTI is located at Palomar Observatory and is equipped with two 40 cm
siderostats separated along a 110 m N--S baseline.\footnote{ PTI can
also be reconfigured for a 85 m N--W baseline, not utilized in this
experiment.} Observations of Vega were made in the K band (2.2 \micron)
on 9 nights over two observing seasons: 1999 May 24, 25 \& November 4,
and 2000 May 09 \& July 1, 4--6, 26. The K filter used is a good match
to the CIT photometric system \citep{colavita99, elias82, elias83} and
is sampled in five spectral channels ($R \sim 20$) centered at
$\lambda_c = 2.009,\ 2.106,\ 2.203,\ 2.299,\ 2.396$ \micron. The
spectral channel photons are sent through an optical fiber, which
behaves like a spatial filter restricting the field of view of the
fringe tracker to 1\arcsec\ (Gaussian FWHM). PTI is, therefore,
effectively insensitive to emission on scales of several arcsec or
larger. The fringe contrast or the squared visibility ($V^2 $) of the
source brightness distribution projected on the sky is the resulting
observable of the interferometer.

Vega, along with calibration sources, was observed $2-5$ times during
each night, and each observation was approximately 130 seconds long. The
calibration of the Vega $V^2$ data, on a channel-to-channel basis, is
performed by estimating the interferometer system visibility
($V_{sys}^2$) using calibration sources with model angular diameters and
then normalizing the raw Vega visibility by $V_{sys}^2$ to estimate the
measured $V^2$ \citep{mozurkewich91, boden98}. Uncertainties in the
system visibility and the calibrated target visibility are inferred from
internal scatter among the data in a scan, the uncertainties associated
with the predicted calibrator diameters, and standard error-propagation
calculations.

The calibrators used are main sequence stars with predicted unresolved
angular sizes of $<0.75$ mas: HD 166620 (K2V), HD 166014 (B9.5V), and HD
168914 (A7V). Calibrating the three calibration objects against each
other produced no evidence of systematic errors, with all objects
delivering reduced $V^2 = 1$. A summary of the calibrators is given in
Table 1.

The data collected on the various nights were sorted by projected
baseline and wavelength and were averaged via a uncertainty weighted
mean. The uncertainties of the means were estimated from the variance of
the input data. The reduced dataset contains a set of four data points
per spectral channel, and is displayed in Figure 1. The apparent
discrete spatial frequency sampling is a direct result of the spectral
sampling of the K-band filter. Vega transits nearly overhead at PTI;
consequently, the projected baseline on the sky changes by only a few
percent. Thus, the spatial frequency sampling ($B/\lambda$) in Figure 1
is dominated by the spectral channels and not by changes in the
projected baseline. Also shown in Figure 1 are separate uncertainty
weighted means for the data collected in 1999 and the data collected in
2000 data. 

In addition to the standard visibility calibrators, Altair was also
observed on the same nights as Vega (only during the 2000 season). The
Altair observations are described in detail by \citet{vB01}. In many
ways, Altair is the near-perfect comparison star for Vega.

Altair is located in the same part of the sky as Vega, making it
observable on the same nights. Altair is of similar spectral type
(A7IV-V) to Vega which helps minimize comparison difficulties that may
arise from differences in the spectral slopes across the broad K band
filter. Altair is of similar brightness (V=0.77, K=0.26 mag) and angular
size (3.4 mas; \citet{vB01}) to Vega. Because the signal-to-noise ratio
of the observations is dependent upon both the brightness and the
angular size of a source, the similar brightnesses and angular sizes
yield similar data quality. Finally, Altair is not known to have an
infrared excess \citep[e.g.,][]{dbr97, kbk98}. Thus, the visibility
curve for Vega can be compared to that of a resolved and uncontaminated
stellar profile of similar spectral type, brightness, and angular size.

Altair is, however, rotating at a substantial fraction of its critical
velocity and is viewed at a high inclination \citep[$i \sim 40^\circ -
60^\circ$; ][]{j72,vB01}. As a result, Altair presents an (on-sky)
elliptical stellar disk. In the analysis here, only the N-S baseline
data for Altair has been utilized, which is relatively insensitive to
the oblateness of the photosphere, and is well represented by a single
angular diameter. A detailed summary of the interferometric data and
analysis for Altair is presented by \citet{vB01}. \citet{vB01} averaged
the data over wavelength, but here, the wavelength information has been
maintained to ensure a proper comparison to Vega.

The Altair data are shown in Figure 2. Because Altair is located at a
declination which is 30\degr\ lower than Vega, the magnitude of the
projected baseline changes more significantly for Altair than for Vega.
Thus, the changing projected baseline, coupled with the spectral
sampling, results in a more uniformly distributed sampling of the
spatial frequencies than what is associated with the Vega data.

\section{Discussion}

\subsection{Apparent Stellar Disk}

The simplest interpretation of the data shown in Figure 1 is to assume that
only the stellar photospheric disk contributes to the observed visibility
function. A source which presents a uniform disk of angular size $\Theta_{UD}
$ will yield a visibility curve of the form: 
\begin{equation}
V^2 = \left[\frac{2J_1(\pi(B/\lambda)\Theta_{UD})} {\pi(B/\lambda)\Theta_{UD}
}\right]^2
\end{equation}
where $J_1$ is the Bessel function of first order, $B$ is the magnitude
of projected baseline vector, $\lambda$ is the wavelength of the
observations, and $\Theta_{UD}$ is the apparent stellar uniform disk
angular diameter. The visibilities were fitted under the assumption that
the structure is independent of wavelength within the K filter.

The best fit angular diameter was found by evaluating equation (1) with
a range of angular diameters ($\Theta_{UD} = 0.1-10$ mas, $\Delta
\Theta_{UD}=0.001$ mas). The $\chi^2$ was calculated for each test
value, and the best fit was determined by minimizing the $\chi^2$. The
uncertainty was estimated via a Monte Carlo simulation where the data
points were randomly adjusted by their individual uncertainties, and the
data were re-fitted. The simulation was performed 5000 times, and the
final uncertainty was estimated from the standard deviation of the best
fit angular diameters. The best fit uniform disk diameter was found to
be $ \Theta_{UD}= 3.24 \pm 0.01$ mas, with a reduced $\chi^2/\nu \sim
2.7$. Based upon the probability distribution of $\chi^2$, there is
$<1$\% chance of exceeding such a large $\chi^2/\nu$.

While limb darkening in A stars at 2.2 \micron\ is expected to be relatively
low \citep[e.g.][]{claret00}, assuming that the star is a simple uniform
disk will cause an underestimation of the true, limb-darkened disk size of
the star. Adapted from \citet{hb74}, the visibility function for a linear
limb darkened stellar disk model can be parameterized as: 
\begin{equation}
V^2 = \left[\frac{1-\mu_{\lambda}}{2} + \frac{\mu_{\lambda}}{2}\right]^{-2}
\left[\frac{(1-\mu_{\lambda})J_1[\pi(B/\lambda)\Theta_{LD}]} {
\pi(B/\lambda)\Theta_{LD}} + \frac{(\mu_{\lambda})j_1[\pi(B/\lambda)
\Theta_{LD}]} {\pi(B/\lambda)\Theta_{LD}}\right]^2
\end{equation}
where $\mu_\lambda$ is the linear limb darkening coefficient ($\mu_{2.2 \mu 
\rm{m}} \approx 0.15$ for Vega; \citet{claret00}), $j_1$ is the
spherical Bessel function of first order, and $\Theta_{LD}$ is the apparent
stellar limb darkened disk angular diameter. \citet{hb74} estimate that a
linear approximation to limb darkening has $<0.5$\% uncertainty in the
determination of the stellar disk size, as long as the data sample the first
lobe of the Bessel function (as our data do). A more complicated limb
darkening model (e.g., a quadratic) could be applied; this would contribute
additional Bessel function terms (both normal and spherical), but at ever
decreasing contribution. The added terms most significantly alter the
visibility function beyond the primary lobe (i.e., past our data sampling).

Evaluating $\Theta_{LD}$ over the range of $0.1-10$ mas ($
\Delta\Theta_{LD}=0.001$ mas), equation (2) was sampled in a similar
manner to equation (1). The best fit limb darkened stellar disk diameter
was determined to be $\Theta_{LD} = 3.28\pm0.01$ mas, with a reduced
$\chi^2/\nu \sim 2.7$, no better than the uniform disk model.

The best uniform disk and limb darkened disk models are shown in Figure
1. On the scale of the plot, the two models are nearly
indistinguishable. The observations sample the visibility curve along
the first lobe of the Bessel functions; however, most of the power
associated with limb darkening occurs near the first null and beyond
(i.e., at higher spatial frequencies; \citet{hb74}). Thus, our data are
relatively insensitive to limb darkening, except as a matter of scaling
\citep[e.g.,][]{hb74}. The apparent angular diameter of the stellar disk
for Vega is underestimated by a factor of $\sim 1.2\%$, when a uniform
disk model is assumed. The angular diameter measurement of Vega
presented here represents the most precise size estimate of Vega to
date. Previously, \citet{hbda74} measured the $\lambda=0.44$ \micron\
angular size of Vega and derived a limb darkened stellar diameter of
$3.24\pm0.07$ mas, which agrees with our measurement within their
uncertainties.

Coupled with knowledge of the bolometric flux of the star, the measured
angular diameter yields the effective temperature: 
\begin{equation}
T_{eff} = 2341\left[\frac{F_{bol}}{\Theta_{R}^2}\right]^{1/4}
\end{equation}
where $F_{bol}$ is the bolometric flux in units of $10^8$ erg cm$^{-2}$
s$ ^{-1}$, and $\Theta_{R}$ is the mean Rosseland (photospheric) angular
diameter in mas. Estimating $\Theta_R$ with the derived $\Theta_{LD}$,
the bolometric flux for Vega ($F_{bol} = 2983\pm120 \times 10^8$ erg
cm$^{-2}$ s$ ^{-1}$; \citet{aam94}) yields an effective temperature of
$T_{eff} = 9553 \pm 111$ K.\footnote{ The uncertainty in the derived
effective temperature is dominated by the uncertainty in the bolometric
flux.} Recent atmosphere models by \citet{ck94} indicate that Vega has a
temperature in the range of $T_{eff} = 9550 - 9650$ K.

While our results are in general agreement with previous work, the
reduced $ \chi^2/\nu \sim 2.7$ (for either of the stellar disk models)
is not very good. The residuals (bottom panel of Figure 1) show that the
uniform disk and limb darkened disk models overestimate the visibility
at lower spatial frequencies and underestimate the visibility at higher
spatial frequencies. If the residuals are a result of the
underestimation of the limb darkening coefficient, increasing its value
should decrease the residuals. In order to significantly improve the
quality of the fit ($\chi^2/\nu \sim 1$), the limb darkening coefficient
would need to be an unrealistic $\mu_{\lambda} \gtrsim 1$. This value is
7 times larger the anticipated near-infrared limb darkening coefficient,
and is twice as large as the optical coefficient. It is unlikely,
therefore, that simply underestimating the limb darkening coefficient is
the source of the residuals.

Over half of the data points are located $>1\sigma$ away from the best
fit models. But perhaps more suggestive is the fact that the residuals
appear linear as a function of spatial frequency. A linear,
uncertainty-weighted, least squares fit was performed to the residuals,
which is shown in the bottom panel of Figure 1. The slope of the
residuals is non-zero at the $ 6\sigma$ level ($slope = 0.0018\pm
0.0003$), suggesting that the residuals shown in Figure 1 are
significant. But from where do they originate?

\subsection{Instrumental Artifacts}

The first test was to determine if the residuals could be an artifact
introduced into the data from either the hardware or the data reduction
process. The 1999 data and the 2000 data were compared to search for
possible systematic errors introduced by the hardware. In Figure 1, the
1999 data and the 2000 data, both averaged as a function of wavelength,
are overplotted to determine if there was a systematic difference
between the two observing seasons. The two data subsets agree extremely
well with each other. In fact, their agreement is even more remarkable
when it is noted that the optics for PTI were disassembled, re-coated,
and re-assembled over the 1999-2000 winter. Given the agreement between
the 1999 and 2000 data, and the fact that PTI is re-aligned nightly, it
is difficult to understand how the residuals could be produced by the
hardware in such a consistent manner.

A second possibility is that the residuals are introduced into the data
during the reduction process. To test for this, the Altair data were
reduced in a manner similar to how the Vega data were reduced. The
Altair data are shown in Figure 2 with a best fit uniform disk of
$\Theta_{UD} = 3.42\pm0.02$ mas, and the residuals of the model fit are
shown in the lower panel of Figure 2. The uniform disk fit for Altair is
very good ($\chi^2/\nu \sim 0.8$ ), and the model residuals do not
appear to display the same organized structure as is seen in the Vega
data. Thus, the same processing techniques did not appear to create the
same type of residuals observed for Vega. Based upon the above analyses,
it is difficult to understand the Vega model residuals in terms of
instrumental effects.

\subsection{Stellar Surface Features}

While stellar spots are not expected on a hot star such as Vega, it is
possible that surface features are present, and the stellar disk would
no longer be illuminated uniformly. The observed visibility curve would
then be altered, in comparison to a uniform disk or limb darkened disk
model. It is conceivable, therefore, that the residuals are a result of
star spots on the surface of Vega.

To test for this, the limb darkened disk plus star spot model developed
by \citet{vB01} was adapted and applied to the Vega data. The model is a
limb darkened disk of diameter 3.28 mas with a randomly placed bright
spot which covers 25\% of the stellar surface. The resulting visibility
curve is shown in Figure 3, and is compared to the data and the 3.28 mas
LD curve from Figure 1.

The spots are on a smaller spatial scale than the stellar photospheric
disk, thus contributing the majority of their power at spatial
frequencies higher than where the data are located. That is, the spots
affect mostly the location of the first null and the amplitude of the
secondary lobes.

Near the location of the data, stellar surface features do affect the
visibility curve on the scale of the residuals. However, as can be seen
in the Figure 3, a single spot merely shifts the visibility curve
without significantly changing the slope of the curve. Thus, a single
spot can not reproduce the visibility curve observed for Vega. It is
possible that a suite of spots in the correct combination could
reproduce the residuals observed in the data.

The good agreement between the 1999 and 2000 data implies that the
stellar surface features on Vega would have to persist over the period
of 7 months or would have to reappear in such a way as to reproduce
nearly the exact same visibility curve. So while, in principle, stellar
surface features could produce the observed visibility curve residuals,
this possibility is viewed as unlikely given the good agreement of the
data separated by over 7 months.

\subsection{Stellar Atmosphere Structure}

As described in \S2., the data spectrally sample the K band filter at 5
different wavelengths (2.009, 2.106, 2.203, 2.299, \& 2.396 \micron,
$\Delta\lambda = 0.096$ \micron), but the $V^2$ data were fitted under
the assumption that the stellar structure is independent of wavelength
across the K filter. In principle, the data may be sampling different
temperatures (layers) within the stellar atmosphere, and this might be
the source of the residuals. To test for this, the limb darkened model
in equation (2) was applied to the data as a function of wavelength.

For each of the five wavelengths within the data, it is assumed that the
limb darkening coefficient is the same ($\mu_{\lambda} = 0.15$), and the
apparent limb darkened stellar disks as a function of wavelength are
derived. To ensure that the assumption of grey opacity across the K
filter is an adequate approximation, an additional experiment was
performed. As a function of wavelength, the limb darkening coefficient
was adjusted until the derived stellar disk equaled (to within the
uncertainties of the fitting) 3.28 mas, (the stellar diameter that best
matched the dataset as a whole). The limb darkening fitting required a
monotonically decreasing limb darkening coefficient with a full range of
$\mu_{2.009 \mu \rm{m}} = 0.20 $ to $\mu_{2.396 \mu \rm{m}} = 0.05$.
However, for a 10,000 K star, a linear limb darkening coefficient of
$\mu_{\lambda} = 0.20$ corresponds to a wavelength between the J (1.25
\micron) and H (1.65 \micron) photometric bands, and $\mu_{\lambda} =
0.05$ corresponds to a wavelength $\lambda \gg 2.2$ \micron\
\citep{cdg95}. Thus, the range of required limb darkening coefficients
appears extreme, and the assumption of grey opacity across the K filter
is deemed adequate for the test of whether atmospheric structure is
being sampled by the interferometer.

The derived wavelength-dependent limb darkened stellar diameters are
shown in Figure 4. At 2.009 \micron, the apparent limb darkened disk
diameter is $\Theta_{2.009 \mu \rm{m}} = 3.24 \pm 0.03$ mas, and at
2.396 \micron, the apparent disk size is $\Theta_{2.396 \mu \rm{m}} =
3.33 \pm 0.04$ mas -- a size increase of $\sim 3$\% across the K band
filter. The overall trend is represented by a linear fit of the form:
\begin{equation}
\Theta_{LD} = (2.8\pm0.2\ mas) + \left(0.21\pm0.01 \frac{mas}{\micron}
\right)\lambda.
\end{equation}

While the angular diameter trend across the K band filter is apparently
linear, the parametric fit fails to predict reasonable
wavelength-dependent limb darkened diameters in two ways. First, the
linear relationship indicated by equation (4) predicts an optical
($\lambda = 0.44$ \micron) limb darkened stellar diameter of 2.9 mas --
$5\sigma$ smaller than the measured limb darkened diameter
\citep[$\lambda=0.44$ \micron, $\Theta_{LD} = 3.24 \pm 0.07 $
mas;][]{hbda74}. The optical size measurement of Vega should be
sensitive to only the stellar photosphere. Dust emission at optical
wavelengths is highly unlikely, and optical polarization studies have
detected no polarization as would be expected for optical scattering
\citep{md98, bm00}. Second, the above linear relationship can not hold
over all wavelengths as it predicts an infinite angular size at infinite
wavelength. It is, of course, possible that because the data sample only
a small range in wavelength, the relationship only appears to be linear,
but it is difficult to reconcile such a linear progression across the K
filter with the good agreement between the measured 0.44 \micron\ and
2.009 \micron\ angular diameters.

Is a 3\% apparent size increase across the K filter even feasible for a
$T_{eff}\approx 9600$ K main sequence star? To answer this question, a
NextGen stellar atmosphere model ($T_{eff}=9600$ K, $\log {g}=4.0$,
[M/H]=0.0) by \citet{hab99} was utilized. By assuming that at any given
wavelength, the stellar atmosphere is probed to an optical depth of
$\tau _{\lambda }=2/3$, the originating temperature layer within the
star can be estimated by equating the observed flux $F_{\lambda }$ to
$(2/3)B_{\lambda }(T)$. Convolving the spectral channels with the
atmosphere model and solving for the temperature in an iterative
fashion, the originating temperature is converted into a stellar radius
via the temperature versus radius relationship of the atmosphere model.
The flux across the K filter is found to arise from stellar radii
located within 0.05\% of each other -- a value significantly smaller
than the 3\% implied by Figure 4 and equation (4).

The above analysis suggests that radial sampling of the stellar
atmosphere across the K filter can not explain the observed visibility
function. It is cautioned, though, that the models by \citet{hab99} are
plane parallel atmosphere models, and Vega is suspected to be fast
rotator ($v\sin i = 245$ km s$^{-1}$), viewed nearly pole-on \citep[$i
\sim 5-6 \degr$ ;][]{gha94, hwk98}, potentially complicating the above
analysis. However, work by \citet{j72} indicates that a fast rotator
viewed pole-on would appear merely as a heavily limb darkened stellar
disk, and would not substantially affect the slope of the visibility
curve at spatial frequencies located within the first null. An
underestimated limb darkening coefficient as the source of the residuals
was already explored in \S3.1, and an unrealistic limb darkening
coefficient ($\mu_\lambda \gtrsim 1$) was required to reproduce the
observed visibility function.

Finally, the Altair data do not appear to show the same residuals that
are displayed in the visibility curve of Vega. As Altair is of slightly
later spectral type (A7IV--V) than Vega (A0V), one might expect that
atmospheric structure would be more evident in Altair than in Vega. This
is apparently not the case indicating that either the residuals are not
a result of atmospheric structure, or the nearly edge-on view of Altair
($i \sim 45\degr - 60\degr$; \citet{j72,vB01}) has hidden the
atmospheric structure.

Sampling of the radial stellar atmospheric structure across the K band
filter appears unlikely as the source of the observed residuals, as this
requires a 3\% increase in the apparent stellar disk across the K filter
alone. Further, the apparent stellar disk versus wavelength relationship
predicts an optical size much smaller than what is observed. However,
the stellar atmospheric structure, rotational velocity, and on-sky
orientation of Vega are not fully understood \citep[e.g.,][]{ck94,
gha94}, and given the limited sample of the visibility function,
atmospheric structure can not be fully excluded as the origin of the
residuals.

\subsection{Debris Disk Contribution}

Current far-infrared and sub-millimeter observations indicate that the
debris disk around Vega extends out to $20\arcsec - 40\arcsec$, and is
apparently viewed nearly face-on \citep{bp93, hwk98, md98}. The disk
contains a total mass of $M_d \sim 0.5\ M_{Moon}$, and has an infrared
luminosity of $L_{IR} \sim few \times 10^{-3} L_\sun$ \citep{bp93}. The
disk has not previously been confirmed at wavelengths shorter $\lambda
\lesssim 20 $ \micron, but as discussed in \S1, stellar atmosphere
models have consistently underestimated the measured near-infrared
($\lambda \gtrsim 2-5$ \micron) flux of Vega. Altair does not have a
detectable infrared excess at any wavelength \citep{cheng92, bp93,
kbk98}, and the residuals in the Vega data are not matched in the Altair
data. Can the debris disk be responsible for the residuals observed in
the 2.2 \micron\ visibility function of Vega?

To test whether the debris disk can alter the stellar disk visibility
function in such a way as to reproduce the observed curve, a simple
model was created. The model consists of a central, limb darkened
($\mu_\lambda$) star of angular diameter $\Theta_{LD}$, surrounded by a
uniform intensity ring viewed face-on. The true disk is, of course, not
likely to be uniformly illuminated, but adding a more elaborate
intensity distribution increases the complexity of the model, which is
already difficult to constrain with the current dataset. Additionally,
more complex geometries introduce structure at higher spatial
frequencies, and the data, located along the first lobe of the Bessel
function, are largely insensitive to higher frequency complexities.

The model ring has an inner radius $r_i$ and an outer radius $r_o$, and
it contributes a fraction $f$ to the total flux of the system. The
uniform ring can be approximated by the subtraction of a smaller disk of
angular diameter $\Theta_i \propto 2r_i$ from a larger concentric disk
of angular diameter $\Theta_o \propto 2r_o$. The resulting visibility
function is simply an addition of the components weighted by their
relative flux contributions:
\begin{eqnarray}
V^2 & = & \left[\frac{(1-f)}{\frac{1-\mu_{\lambda}}{2} + \frac{\mu_{\lambda}
}{2}} \left(\frac{(1-\mu_{\lambda})J_1[\pi(B/\lambda)\Theta_{LD}]} {
\pi(B/\lambda)\Theta_{LD}} + \frac{(\mu_{\lambda})j_1[\pi(B/\lambda)
\Theta_{LD}]} {\pi(B/\lambda)\Theta_{LD}}\right)\right. \ \rm{(Stellar\
Disk)}  \nonumber \\
& + & \left.2f\frac{J_1(\pi(B/\lambda)\Theta_{o})} {\pi(B/\lambda)\Theta_{o}}
-2f\frac{J_1(\pi(B/\lambda)\Theta_{i})} {\pi(B/\lambda)\Theta_{i}}\right]
^{2} \ \rm{(Uniform\ Ring)}.
\end{eqnarray}

The outer radius ($r_{o}$) of the model disk was set to $0\farcs5$ (4
AU), the HWHP radius of the field of view of the fringe tracker. The
true disk extends well beyond this point, but PTI is insensitive to such
extended emission. If the disk contributes to the visibility function,
as measured by PTI, the disk \emph{must} extend to within $\sim 4$ AU
of the star. Modeling of the longer wavelength infrared excess ($\lambda
\gtrsim 10$\micron) does not require the presence of dust grains located
so close to the star, but the observations also do not exclude such a
possibility \citep{aumann84, bp93}. The inner radius ($r_{i}$) of the
model ring was set by the radius at which the dust grains are expected
to sublimate ($T_{dust}\sim 1500$ K). Assuming pure blackbody grains,
the sublimation radius of the grains can be estimated by 
\begin{equation}
r\approx L^{0.5}\left( \frac{278}{T_{dust}}\right) ^{2}\ \rm{AU}
\end{equation}
where $r$ is in AU, $T_{dust}$ is in K, and $L$ is in $L_{\sun}$. For
Vega, $ L\approx 54\ L_{\sun}$ which sets the inner radius of the disk
to $r_{i}\sim 0.25$ AU ($\Theta _{i}\sim 64$ mas).

The stellar disk ($\Theta_{LD}$) was constrained by the temperature
range of current atmosphere models for the stellar photosphere of Vega
and equation (3). The highest estimate of the effective temperature for
Vega is $T_{eff} \lesssim 9700$ K \citep{db80}, which corresponds to a
stellar diameter limit of $\Theta_{LD} \gtrsim 3.19$ mas. Apart from
this lower limit restriction on the stellar size, the stellar disk
diameter ($\Theta_{LD}$) and the fractional contribution ($f$) of the
disk (the only free parameters allowed) were adjusted
($\Delta\Theta_{LD} = 0.01$ mas, $\Delta f = 0.5$\%) until the $\chi^2$
was minimized.

The best match to data was found with a stellar diameter of $\Theta_{LD}
= 3.20$ mas and a disk contribution of $f=5$\%. The resulting visibility
function is shown in Figure 5 where it is compared to the data and the
$\Theta_{LD} = 3.28$ mas limb darkened stellar disk. The resulting fit
has a reduced $\chi^2/\nu \sim 1.4$ (LD disk model, $\chi^2/\nu \sim
2.7$). An estimate of the uncertainties associated with the model
parameters was made by finding the models which yielded a $\chi^2/\nu
\sim 1.7$. Based upon the probability distribution of $\chi^2$ and the
number of degrees of freedom, there is approximately a 5\% chance of
exceeding such a $\chi^2/\nu$. The two ``bounding'' models have
parameters of $[f=3\%, \Theta_{LD} = 3.22\ \rm{mas}]$ and $[f=6\%,
\Theta_{LD} = 3.2\ \rm{mas}]$. These models are displayed in Figure 5.

While the star+ring model does not match perfectly to the data (and
given the simplicity of the model, a perfect match was not anticipated),
a small 5\% flux contribution of the uniform ring has adjusted the slope
of the visibility curve in precisely the manner required to reduce the
residuals. In contrast to the limb darkened star model, the star+ring
model does not overestimate and underestimate the visibility at lower
and higher spatial frequencies.

With the disk parameters ($\Theta_i$ \& $\Theta_o$) fixed, the stellar
diameter and the disk flux contribution are fairly well constrained. The
disk radii were increased and decreased by a factor of 100 with no
significant change in the results. Even if the inner radius of the disk
is allowed to be at the stellar limb ($\sim 3.2$ mas), a minimum disk
contribution of 2-3\% is always needed to account for the observed
residuals. If all of the parameters are adjusted simultaneously
($\Theta_{LD},\ \Theta_i,\ \Theta_o,\ f$), a ``near perfect'' solution
can be found, but all of the parameters play against each other, and a
single \textit{unique} solution is not found. Interestingly, the
star+ring model appears to be most sensitive to the flux contribution
$f$ of the disk rather than the radii of the ring. This is likely an
indication that the disk edges are ``outside'' the normal size
sensitivity of PTI ($\sim 4$ mas), but still detectable by the
interferometer.

Is this simplistic model realistic? The derived stellar diameter of 3.2
mas is in reasonable agree with the optical limb darkened diameter
$\Theta_{LD}=3.24 \pm 0.07$ \citep{hbda74}. As mentioned above, the
optical size measurement of Vega should be sensitive to only the stellar
photosphere. Using equation (3), the implied effective temperature for
an angular diameter of 3.2 mas is $T_{eff} = 9670$ K, within the range
of currently accepted values for the effective temperature of Vega.
Thus, the model yields a reasonable stellar diameter.

The $f=5$\% flux contribution of the debris disk is approximately a
factor of two (2) larger than the excess predicted by atmosphere models
\citep[e.g.,][]{mountain85}, and a factor of 1.5 larger than the excess
deduced by \citet{megessier95}. Thus, a 5\% flux contribution, while not
an outrageous requirement, is marginally larger than what is expected
from current observations and models. However, the $f=3$\% lower limit
for the infrared excess is certainly within the current photometric
uncertainties of Vega. How much dust mass is required to produce a
$3-6$\% infrared excess at 2.2 \micron?

If it is assumed that 1) the grains emit as blackbodies, 2) the grains
are at a single equilibrium temperature ($T_{dust}$), 3) the grains are
spherical and the size distribution can be approximated with a single
size ($a \sim 10-20$ \micron), and 4) the average grain density is
$\rho_{dust} \sim 3$ g cm$^{-3}$, the dust mass needed to produce a
$3-6$\% 2.2 \micron\ excess can be estimated. A (upper limit) dust
temperature of $T_{dust} \sim 1500$ K implies a necessary dust mass of
only $M_{dust} \sim 10^{-5}\ M_{Moon}$. If a lower temperature of
$500-700$ K is used \citep[e.g.,][]{ssb97}, a dust mass of $M_{dust}
\sim 10^{-3}\ M_{Moon}$ is required. \citet{aumann84} suggested that as
much as 10$^{-3}$ of the emitting grain area (which is proportional to
the grain mass for a single grain density and size) could be at 500 K
and not violate the limits set by the {\it IRAS} observations. Thus, a
rough estimate of the dust mass necessary to produce the inferred 2.2
\micron\ excess is within the limits of current models for the debris
disk.

It is cautioned that the model of a stellar disk surrounded by a uniform
ring is not intended to imply that the disk surrounding Vega is a
uniform ring of material. The model was created to test whether extended
structure around the central star could reasonably explain the observed
residuals in the visibility function of Vega. More complete spatial
frequency sampling is necessary to constrain a more complex (realistic)
disk model. The presented results imply that the debris disk is a viable
explanation.

\section{Summary}

Near-infrared (2.2 \micron) interferometric observations of Vega have
been obtained. The resulting visibility data have been modeled with a
$3.28 \pm 0.01$ mas diameter limb darkened stellar disk. After
subtraction of the stellar disk model, residuals within the data appear
organized and significant. Four possible sources of the residuals are
discussed in detailed: instrumental artifacts, stellar surface features,
stellar atmospheric structure, and the surrounding debris disk.

Emission/scattering from the debris disk is found to be the most likely
explanation. A simple star + uniform ring model was developed to
simulate the effect of the debris disk upon the visibility function. A
$3-6$\% flux contribution from dust located within $r \lesssim 4$ AU can
alter the stellar disk visibility function in precisely the manner that
is observed. The interferometric observations appear to be mostly
sensitive to the relative flux contribution of the disk and not to the
geometry of the disk. If the source of the disk contribution is emission
from warm dust, a (blackbody) dust mass of $\sim 10^{-3}\ M_{Moon}$ at
$T_{dust} \approx 500-700$ K could account for the apparent
contribution.

Our findings are very intriguing but inconclusive. Infrared polarization
and/or infrared interferometric observations, which have better
$uv$-plane coverage than those obtained with PTI, are required to verify
and explain the results obtained here. The infrared capabilities of
interferometers currently under construction (e.g., Keck Interferometer,
CHARA, or the VLTI) would be invaluable to this project. If the debris
disk has indeed been detected at 2.2 \micron, this has important
consequences towards our understanding of Vega, its debris disk, and
debris disks in general.

\acknowledgments

DRC would like to thank Peter Hauschildt for very useful discussions
regarding the structure of stellar atmospheres, and for kindly providing
the NextGen stellar atmosphere models to the public. DRC was supported
in part by NASA WIRE ADP NAG5-6751, and EAL acknowledges support from a
Research Corporation Innovation Award and a Presidential Early Career
Award for Scientists and Engineers to the University of Florida (NSF
Grant \# 97-33367). Portions of this work were performed at the Jet
Propulsion Laboratory, California Institute of Technology under contract
with the National Aeronautics and Space Administration. The referee is
thanked for adding to the clarity of this paper. A hearty thank you goes
to the entire PTI collaboration for their efforts in maintaining and
running a great interferometer
\url{(http://huey.jpl.nasa.gov/palomar/ptimembers.html)}. DRC and GvB
would like to warn all PTI observers to be on the alert for giant
inflatable bumblebees at 3 in the morning.

\begin{figure}[tbp]
\epsscale{0.75} \plotone{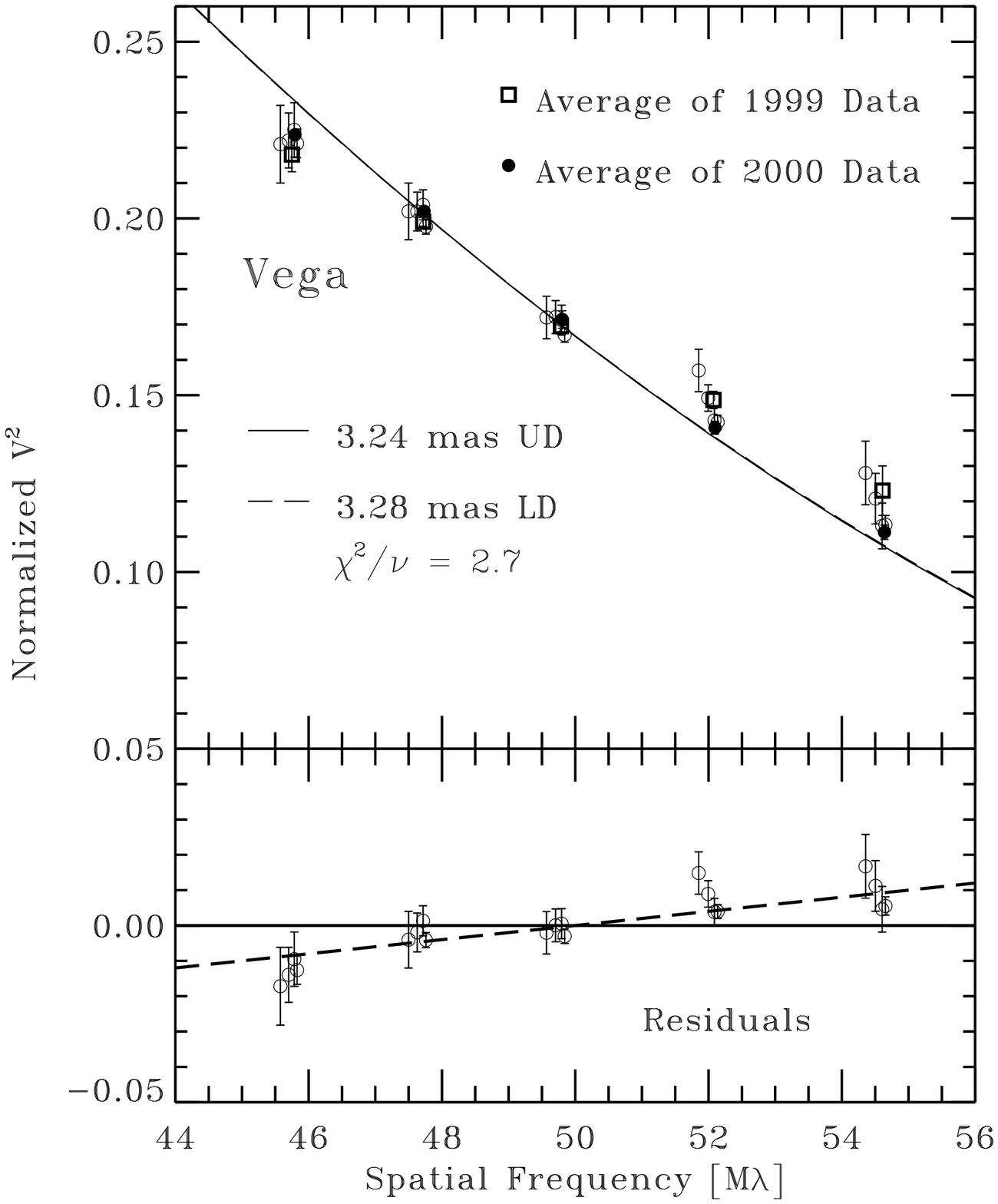} 
\figcaption{{\it Top:} Normalized visibility curve for Vega. The {\it
open circles} represent data averaged as a function of baseline,
regardless of year of acquisition. The 1999 data {\it(open squares)} and
the 2000 data {\it(filled circles)} data, averaged by wavelength are
shown. The $\Theta_{UD}=3.24$ mas uniform disk and the
$\Theta_{LD}=3.28$ mas limb darkened disk models are overplotted and are
nearly indistinguishable. {\it Bottom:} The residuals of the limb
darkened disk model for the baseline averaged data are shown. The
horizontal line marks the zero level, and the {\it dashed} line is a
linear fit to the residuals.}
\end{figure}

\begin{figure}[tbp]
\plotone{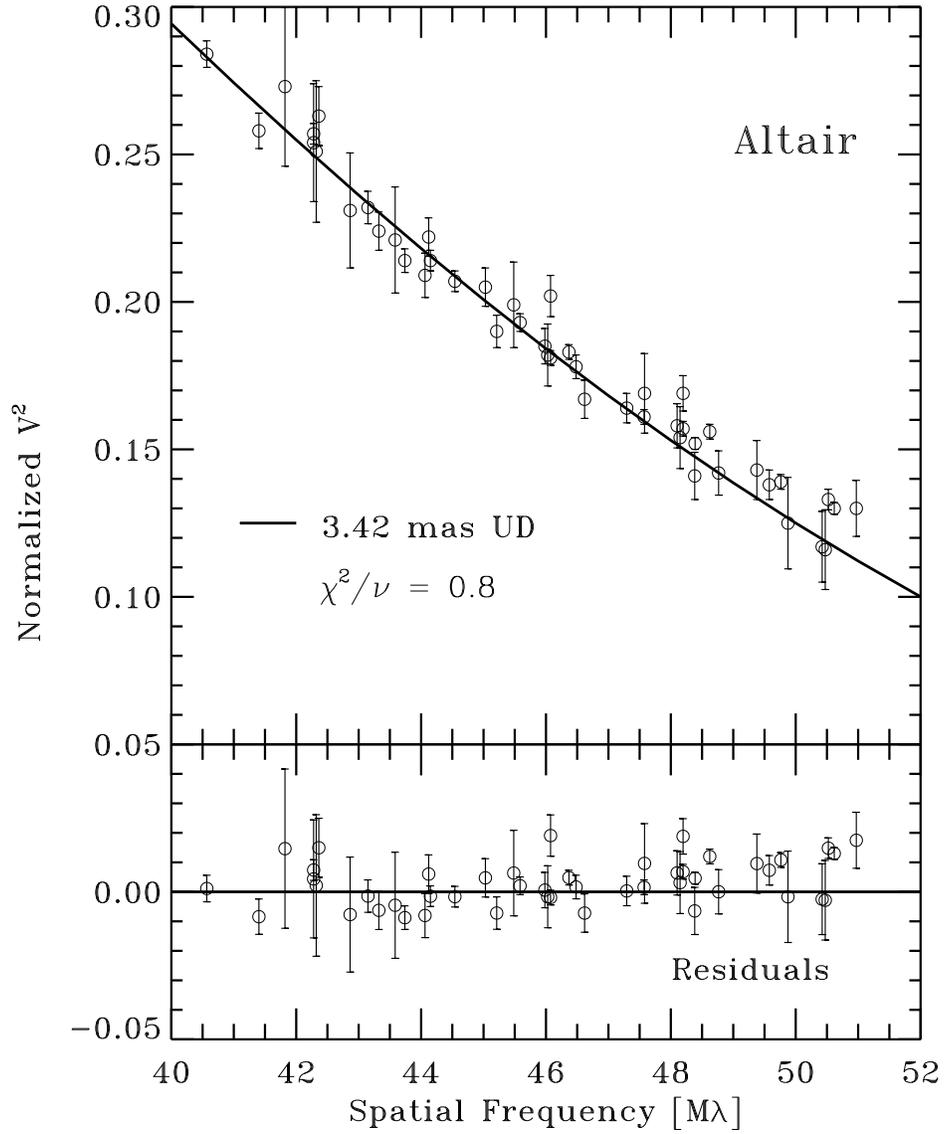} 
\figcaption{{\it Top:} Normalized visibility curve for Altair. The {\it
open circles} represent data averaged as a function of baseline as was
done for Vega (Figure 1). The $\Theta_{UD}=3.42$ mas uniform disk model
is overplotted. {\it Bottom:} The residuals of the uniform disk model are
shown. The horizontal line marks the zero level.}
\end{figure}

\begin{figure}[tbp]
\plotone{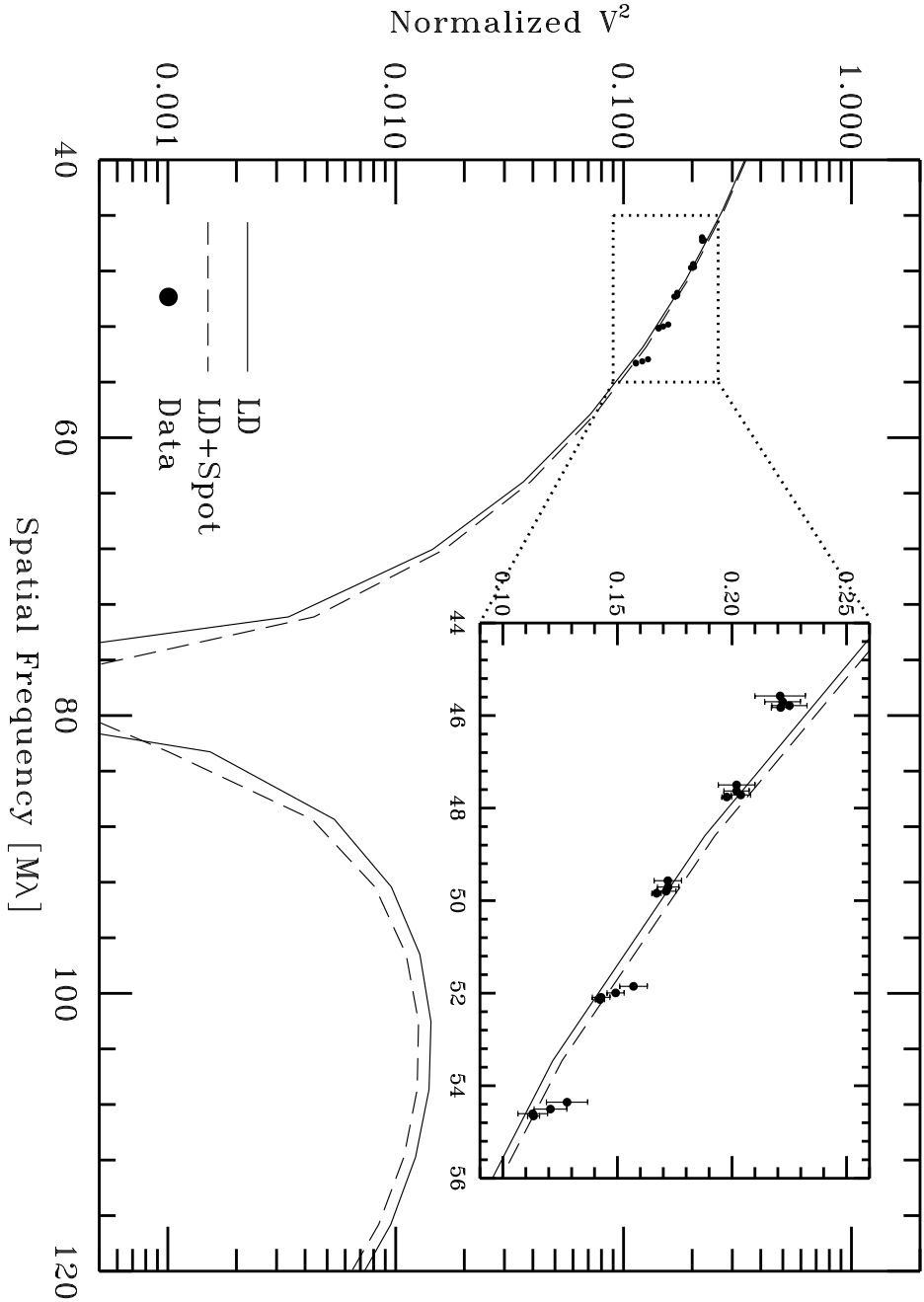}
 
\figcaption{Normalized linear--log visibility curve for Vega. The {\it
solid curve} represents the $\Theta_{LD}=3.28$ mas limb darkened model
from Figure 1. The {\it dashed curve} represents the same limb darkened
model, but with a surface spot covering 25\% of the stellar disk. 
{\it Inset:} Detailed version of the plot with scaling set to that of the
visibility curve shown in Figure 1.}

\end{figure}

\begin{figure}[tbp]
\plotone{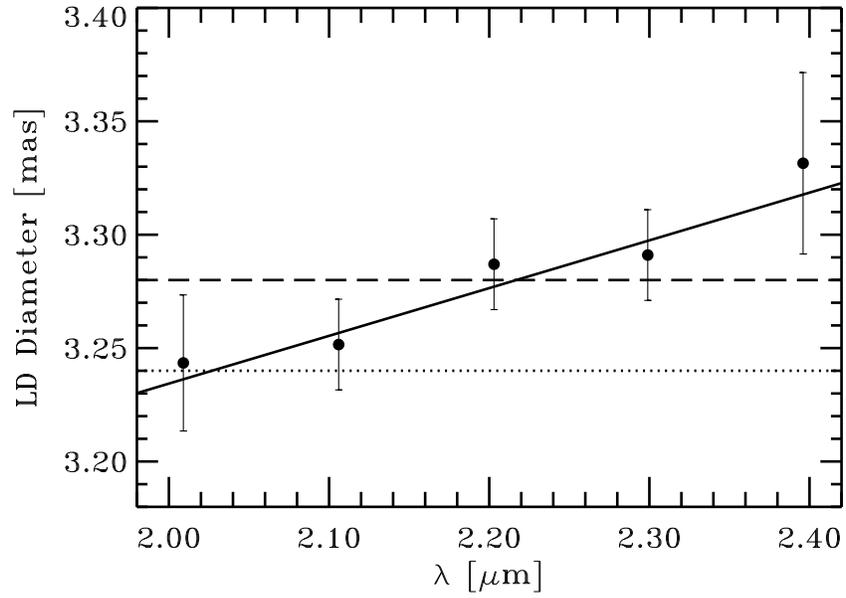}
 \figcaption{Apparent limb darkened disk sizes for Vega shown as a
function of wavelength. The {\it dashed line} represents the angular
diameter of the limb darkened model in Figure 1
($\Theta_{LD}=3.28\pm0.01$ mas), and the {\it dotted line} represents
the limb darkened angular diameter obtained at $\lambda = 0.44$ \micron\
($\Theta_{LD}=3.24\pm0.07$ mas; \citet{hbda74}).}
\end{figure}

\begin{figure}[tbp]
\plotone{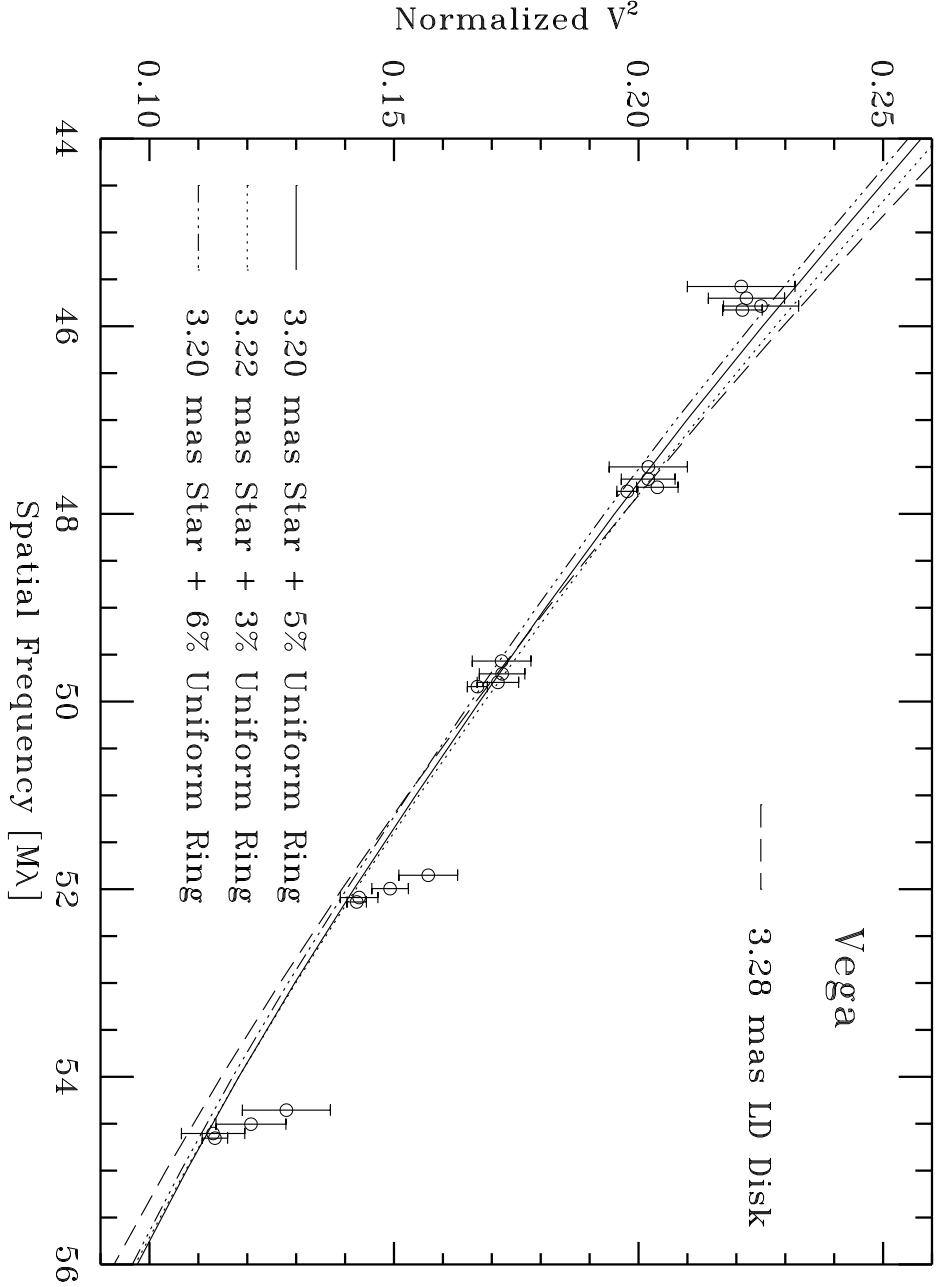} 
\figcaption{Normalized visibility curve for Vega. The {\it dashed curve}
is the stellar limb darkened model from Figure 1. The {\it solid curve}
represents the stellar disk $+$ 5\% uniform ring model ($\chi^2/\nu \sim
1.4$). The {\it dotted curve} and the {\it dash-dot curve} represent estimated
limits to the star $+$ uniform ring models ($\chi^2/\nu \sim 1.7$).}
\end{figure}

\clearpage
\begin{deluxetable}{ccccl}
\tablecolumns{5} \tablewidth{0pc} \tablecaption{Calibration
Sources.} \tablehead{ \colhead{Source} & \colhead{$\theta_{EST}$\tablenotemark{a}}
& \colhead{Distance from}
& \colhead{Spectral}& \colhead{Notes} \\
\colhead{} & \colhead{(mas)} & \colhead{Vega (deg)} &
\colhead{Type} & \colhead{} } \startdata
HD166620 & $0.74 \pm 0.08$ & 5.3 & K2V & Primary calibrator\\
HD166014 & $0.66 \pm 0.10$ & 11.7 & B9.5V & \\
HD168914 & $0.41 \pm 0.20$ & 10.4 & A7V & \\
\enddata
\tablenotetext{a}{The estimated angular diameters come from photometric
fits.}
\end{deluxetable}


\begin{thebibliography}{}

\bibitem[Alonso, Arribas, \& Mart\'{i}nez-Roger(1994)]{aam94}  Alonso, A.,
Arribas, S., \& Mart\'{i}nez-Roger, C. 1994, \aap, 282, 684

\bibitem[Aumann et al.(1984)]{aumann84}  Aumann, H. H. et al. 1984, \apjl,
278, L23

\bibitem[Backman \& Paresce(1993)]{bp93}  Backman, D. E. \& Paresce, F.
1993, in Protostars and Planets III, ed. E. H. Levy \& Lunine, J. I.,
University of Arizona Press, Tucson \& London, 1253

\bibitem[Bhatt \& Manoj(2000)]{bm00}  Bhatt, H. C., \& Manoj, P. 2000, \aap,
362, 978

\bibitem[Bessell, Castelli, \& Plez(1998)]{bcp98}  Bessell, M. S., Castelli,
F., \& Plez, B. 1998, \aap, 333, 231

\bibitem[Boden et al.(1998)]{boden98}  Boden, A.F, et al., 1998, \apjl, 504,
L39

\bibitem[Bohlin et al.(1990)]{bohlin90}  Bohlin, R. C., Harris, A. W., Holm,
A. V., \& Gry, C. 1990, \apjs, 73, 413

\bibitem[Castelli \& Kurucz(1994)]{ck94}  Castelli, F. \& Kurucz, R. L.
1994, \aap, 281, 817

\bibitem[Cheng et al.(1992)]{cheng92}  Cheng, K.-P., Bruhweiler, F. C.,
Kondo, Y., \& Grady, C. A. 1992, \apj, 396, L83

\bibitem[Claret, D\'{i}az-Cordov\'{e}s, \& Gim\'{e}nez(1995)]{cdg95}
Claret, A., D\'{i}az-Cordov\'{e}s, J., \& Gim\'{e}nez, A. 1995, \aap,
114, 247

\bibitem[Claret(2000)]{claret00}  Claret, A. 2000, \aap, 363, 1081

\bibitem[Colavita(1999)]{colavita99}  Colavita, M. M. et al. 1999, \apj,
510, 505

\bibitem[Cohen et al.(1992)]{cohen92}  Cohen, M., Walker, R. G., Barlow, M.
J., \& Deacon, J. R. 1992, \aj, 104, 1650

\bibitem[Dreiling \& Bell(1980)]{db80}  Dreiling, L. A. \& Bell, R. A. 1980,
\apj, 241, 736

\bibitem[Dunkin, Barlow, \& Ryan(1997)]{dbr97}  Dunkin, S. K., Barlow, M.
J., \& Ryan, S. G. 1997, \mnras, 286, 604

\bibitem[Elias et al.(1982)]{elias82}  Elias, J., Frogel, J., Matthews, K.,
Neugebauer, G., 1982, \aj, 87, 1029

\bibitem[Elias et al.(1983)]{elias83}  Elias, J., Frogel, J., Hyland, A.,
Jones, T., 1983, \aj, 88, 1027

\bibitem[Gulliver, Hill, \& Adelman(1994)]{gha94}  Gulliver, A. F., Hill,
G., \& Adelman, S. J. 1994, \apj, 429, L81

\bibitem[Hanbury Brown, Davis, \& Allen(1974)]{hbda74}  Hanbury Brown, R.,
Davis, J., \& Allen, L. R. 1974, \mnras, 167, 121

\bibitem[Hanbury Brown et al.(1974)]{hb74}  Hanbury Brown, R., Davis, J.,
Lake, R. J. W., \& Thompson, R. J. 1974, \mnras, 167, 475

\bibitem[Hauschildt, Allard, \& Baron(1999)]{hab99}  Hauschildt, P. H.,
Allard, F., \& Baron, E. 1999, \apj, 512, 377

\bibitem[Hayes(1985)]{hayes85}  Hayes, D. S. 1985, in Calibration of
Fundamental Stellar Quantities, ed. D. S. Hayes, L. E. Passinetti, \& G. D.
Phillip, Dordrecht, Reidel, 225

\bibitem[Heinrichsen, Walker, \& Klass(1998)]{hwk98}  Heinrichsen, I.,
Walker, H. J., \& Klass, U. 1998, \mnras, 293, L78

\bibitem[Jordahl(1972)]{j72}  Jordahl, P. R. 1972, Ph.D. Thesis, University
of Texas at Austin

\bibitem[Kuchner, Brown, \& Koresko(1998)]{kbk98}  Kuchner, M. J., Brown, M.
E., \& Koresko, C. D. 1998, \pasp, 110, 1336

\bibitem[Kurucz(1979)]{kurucz79}  Kurucz, R. L. 1979, \apjs, 40, 1

\bibitem[Leggett et al.(1986a)]{leggett86a}  Leggett, S. K. et al. 1986a,
\aap, 159, 217

\bibitem[Leggett et al.(1986b)]{leggett86b}  Leggett, S. K. et al. 1986b,
\mnras, 223, 443

\bibitem[Mauron \& Dole(1998)]{md98}  Mauron, N. \& Dole, H. 1998, \aap,
337, 808

\bibitem[M\'{e}gessier(1995)]{megessier95}  M\'{e}gessier, C. 1995, \aap,
296, 771

\bibitem[Mountain et al.(1985)]{mountain85}  Mountain, C. M., Leggett, S.
K., Selby, M. J., Blackwell, D. E., \& Petford, A. D. 1985, \aap, 151, 399

\bibitem[Mozurkewich et al.(1991)]{mozurkewich91}  Mozurkewich, D., et al.,
1991, \aj, 101, 2207

\bibitem[Sylvester, Skinner, \& Barlow(1997)]{ssb97}  Sylvester, R. J.,
Skinner, C. J., \& Barlow, M. J. 1997, \mnras, 289, 831

\bibitem[van Belle et al.(2001)]{vB01}  van Belle, G. T., Ciardi, D. R.,
Thompson, R. R., Akeson, R. L. \& Lada, E. A. 2001, \apj, in press
\end{thebibliography}
\end{document}